On Wave Particle Duality

Jerome Blackman
Dept. of Mathematics (Emeritus)
Syracuse University

Introduction. In a recent paper [1] a mathematical model for quantum measurement was presented. The phenomenon of wave particle duality, which is introduced in every beginning course of quantum theory, can be explained using this model. Although it is a special case of the general theory it has historic interest and involves the use of a continuous spectrum which was treated only casually in [1]. The treatment here is self contained except for the proof of a mathematical theorem appearing in [1].

The usual experiment used to demonstrate wave particle duality is the one or two slit experiment. The part of the experiment that is of interest here is what happens to the wave after passing through the slits and while it is interacting with the detector. Imagine an array of n-1 detectors arranged in the x-y plane and suppose the wave function $\phi(x,y,z,t)$ is, for z<0, equal to 0 and is moving parallel to the z axis toward the x-y plane as would be the case if it had already passed through a slit on the way to the detectors. In our model the motion toward the x-y plane will be described as an interaction with the detectors. For simplicity we assume the particle is scalar so that $\phi$ is a scalar function.

The development of the system is assumed to be governed by a one parameter group of unitary transformations starting at t = 0 and ending at t = T. The group is generated by a Hamiltonian but while we assume the initial state of the particle is $\phi(x,y,z,0)$ we can not specify the initial state of the measuring apparatus since it depends on the state of a very large number of particles making up its structure. We can suppose that these various states are labeled by an index set $\Lambda$ so that each $\lambda \in \Lambda$ designates a different initial state and hence also a different path of the state vector in the appropriate Hilbert sphere during the course of the measurement.

Divide the x-y plane into disjoint sets $A_i$ so that for $1 \leq i \leq n-1$ a particle arriving in $A_i$ will cause the $i^{th}$ detector to respond and set $A_n = E_2 - \bigcup_1^{n-1} A_i$ where $E_2$ is the x-y plane. Let $R_i$ be the set of points (x,y,z) such that $(x,y) \in A_i$ and $z > 0$. Let $R^+$ be the set of (x,y,z) with z>0. By "particle" we mean a classical particle if we are using the language of the laboratory or, if we revert to the language of quantum theory, a wave function which is zero outside $R_i$ as it impinges on the detector. The confluence of these two concepts is discussed below but there is a certain blurring of these two ideas when we have to pass between the concepts of the laboratory and that of the quantum formalism.

The program is to first analyze restrictions on the Hamiltonian that governs the time development of the measurement. Since the measurement consists of moving the particle towards the measuring device so that the motion is perpendicular to the x-y plane this is not difficult. The implication of that analysis will show that for each path there is a tendency for the wave function to contract to the condition described in the last paragraph i.e. to be concentrated in a single $R_i$. Finally a measure can be introduced into the space of paths ($\Lambda$) so that this occurs with probability 1 and that the probability of the path ending in $A_i$ is $\iiint_{R_i} |\varphi(x,y,z)|^2 \, dxdydz$ where $\phi(x,y,z) = \phi(x,y,z,0)$.

We will need various Hilbert spaces and some of their subspaces. Let $\mathcal{H}_p$ be the state space of the particle and $\mathcal{H}_m$ the state space of the photo detector array so that $\mathcal{H} = \mathcal{H}_m \otimes \mathcal{H}_p$ is the state space for the particle interacting with the detectors. Let $|x,y,z\rangle$ be the three dimensional Dirac delta function.

Let $\mathcal{H}_i$ be the Hilbert space spanned by the set of all $|x,y,z\rangle$ where $(x,y,z) \in R_i$. An equivalent definition for $\mathcal{H}_i$ is the set of all functions which are square integrable on $R_i$ and which are zero outside of $R_i$. Then $\mathcal{H}_p$ is the direct sum of the $\mathcal{H}_i$ for i = 1, 2,..n. Let $\Omega$ be the set of all subsets of the integers 1, 2,…n and let $w \in \Omega$. Then define $\mathcal{H}_w$ to be the direct sum of the $\mathcal{H}_k$ for all $k \in w$. For example if $w = \{1,3,5\}$ then $\mathcal{H}_w = \mathcal{H}_1 \oplus \mathcal{H}_3 \oplus \mathcal{H}_5$. It is important to note here that the $\mathcal{H}_i$ are mutually orthogonal subspaces.

At this point we can deduce the conditions that are required of the Hamiltonian H that describes the measurement process. It is clear that the movement of the particle parallel to the z axis, imposed by the experimenter, is suggested by the fact that a classical particle starting at (x,y,z) with this motion will describe a motion of the form (x,y,z(t)) ending at (x,y,0). If we think of a classical particle as a quantum particle in an eigenstate of the position operator we would specify that if it is in the state $|x,y,z_0\rangle$ initially then its subsequent states should be of the form $|x,y,z(t)\rangle$ where z(T) = 0 if the measurement ends at time T. In the laboratory a requirement of this precision for each particle at point (x,y,z) is much less realistic than the requirement that if $(x,y,z) \in R_i$ then the point should end up in $A_i$.

The first requirement for the Hamiltonian H is:

(1) H must act in such a way that the final state of the measuring device must be unique for each final state $|p_i\rangle$ of the particle within the accuracy of the experiment.

If the final position is measure by the detectors as above this is clearly satisfied.

Such experiments are designed not only by the requirement that it takes a classical particle into $A_i$ but that it always remain in $R_i$ for $0<t<T$. Since H is the generator of the group this leads to the requirement on H that

(2a) $\qquad H(\mathcal{H}_m \otimes \mathcal{H}_i) \subseteq (\mathcal{H}_m \otimes \mathcal{H}_i)$ for each i.

In [1] it is shown that in the case of a discrete spectrum, which would correspond to each of the sets $A_i$ consisting of a single point, this equation is equivalent to each of the following two statements:

(2b) $\qquad H(\mathcal{H}_m \otimes \mathcal{H}_w) \subseteq (\mathcal{H}_m \otimes \mathcal{H}_w)$ for any $w \in \Omega$ and

(2c) $\qquad$ H is of the form $H = H_1 \otimes I_1 \oplus H_2 \otimes I_2 \ldots \oplus H_n \otimes I_n$

where $I_i$ is the identity on $\mathcal{H}_i$ and $H_i$ is a Hermitian operator on $\mathcal{H}_m$. The proof and meaning of these statements depends on the fact that $\mathcal{H}$ is the direct sum of its subspaces $\mathcal{H}_m \otimes \mathcal{H}_i$:

(3) $\qquad \mathcal{H} = \mathcal{H}_m \otimes \mathcal{H}_1 \oplus \mathcal{H}_m \otimes \mathcal{H}_2 \oplus \ldots \oplus \mathcal{H}_m \otimes \mathcal{H}_n$

and therefore the proof is the same in both cases. This equation also clarifies the meaning of (2c).

From a mathematical point of view (2c) is interesting since it shows the structure of the most general self adjoint operator H which satisfies (2a) but it is likely that the only case of interest to the physicist is the case where all the $H_i$ are the same, say $\bar{H}$. In that case H is of the form

(4) $\qquad\qquad \bar{H} \otimes I_p$

where $I_p$ is the identity on $\mathcal{H}_p$. This reemphasizes the point that the measuring device is designed so that it doesn't have any influence on the states of the particle.

Suppose we start the measurement with the measuring equipment in the state $|g\rangle$ in $\mathcal{H}_m$ and the particle in state $\phi(x,y,z)$ in $\mathcal{H}_p$. Here we are abusing the notation we have been using in the interest of simplicity and common usage. As the measurement proceeds the state will move on the unit sphere of $\mathcal{H}$ according to the unitary group generated by H and the initial state of the apparatus $\lambda$. By (3) we can write the curve in the form

$$(5) \quad \oplus_1^n |g_{\lambda,i}(t)\rangle \otimes \phi_i(x,y,z,t)$$

where $|g_{\lambda,i}(t)\rangle$ is in $\mathcal{H}_m$ and $\phi_i$ is in $\mathcal{H}_i$ and the summation is in $\mathcal{H}$.

If at some time $t_0$ the $i^{th}$ term vanishes then the group property of the unitary group and condition (2a) implies that it will vanish at all subsequent times. Of course, because it must be on the unit sphere it can only reduce to one term. The reason we might expect a term to vanish is that $g_{\lambda,i}(t)$ terms represent the state of the measuring device which is governed by a large number of particles and fields and therefore might be expected to have rapid variations. Our model will reflect that situation.

Since we are only interested in the absolute value of the coefficients we will map the curve (5) into a curve in $R_n$ by the map

$$\sum_{i=1}^n g_{\lambda,i}(t) \otimes \phi_i(x,y,z,t) \dashrightarrow (a_{\lambda,1}(t), a_{\lambda,2}(t), \ldots a_{\lambda,n}(t)) \text{ where}$$

$$(6) \quad a_{\lambda,i}(t) = \| g_{\lambda,i}(t) \otimes \phi_i(x,y,z,t) \|^2.$$

The $a_{i,\lambda}(t)$ are all non negative and $\sum a_{i,\lambda}(t) = 1$. Therefore the curves all lie on the simplex in $R_n$ defined by $\sum_1^n x_i = 1$ with $x_i \geq 0$. In $R_n$ the simplex is of dimension n-1. For n = 3 it is the triangle bounded by the points (1,0,0), (0,1,0) and (0,0,1).

At t = 0 when there is no entanglement (5) has the form
$$|g_\lambda(0)\rangle \otimes \oplus_1^n \phi_i(x,y,z,0).$$

Here $|g_\lambda(0)\rangle$ is the initial state of the apparatus and has norm 1. $\phi_i$ is in $\mathcal{H}_i$ which consists of the square integrable functions on $R_i$. Putting this together with (6) we se that the starting point is given by

$$(7) \quad a_{\lambda,i}(0) = \int_{R_i} |\phi(x,y,z,0)|^2 \, dxdydz$$

where $\phi$ is defined by $\phi = \oplus_1^n \phi_i$ and represents the initial state of the particle. In (7) we see that the initial condition does not depend on $\lambda$ so it can be dropped and the remaining problem is to introduce a probability measure in the set of paths $\Lambda$ so that the probability of the path ending at the point $e_k = (0, 0 ... \delta_k^i, 0...0)$ is given by $a_i(0)$ in equation (7). This problem was solved in [1] and the interested reader can find it there. In brief what is done is to approximate the simplex by a set of discrete points and the time interval by a finite set of intervals. Then the path is treated as a Markov process or a random walk. It is more complicated than a random walk in Euclidean space because of the need to stay on the simplex and there are actually a whole class of probability distributions that lead to the same result

Comments. If the theoretical model presented here and in [1] is accepted by the physics community it would represent a big change in the way a fundamental part of quantum theory is viewed. From the early days of quantum theory there appeared to be two parts in the time development of quantum phenomena. One was the usual evolution through the action of a unitary group which implied reversibility and the other was through the action of a measurement which precluded reversibility. For example in [2] Leonard Susskind describes the evolution of consensus on the question of whether the information was recoverable after matter dropped into a black hole. The result was agreement that it was recoverable even then but still not if the matter was measured in a laboratory. Our present work shows that even a measurement is described by a unitary transformation at least if one could keep track of the particular $\lambda$ that was used in the measurement.

Roger Penrose [3] discusses the question of unitary development versus measurement and conjectures that the fact that while the Schrödinger equation seems to imply that every particle would eventually be entangled with every other particle, the fact that that does not seem to be the case may be due to the measurement operation being ubiquitous in nature. Since a particle which is in an eigenstate of any complete observable will not be entangled that conjecture is quite plausible. It would require an entangled particle to spend an interval of time in a situation where that observable was not changed (as in the case of the x and y coordinates in the case discussed above) while interacting with the local environment in a random fashion. It might also shed light on the difficulty of maintaining coherent states in the field of quantum computing.

It might be noted that the treatment above fits in neatly with the fact that the particle duality phenomenon occurs at all levels of sensitivity of the detector. As the mesh of the detector is made finer and finer all that is required is that the precision of the measurement maintain the validity of equation (2a).